\documentclass[12pt]{iopart}

\newcommand{\Trr}{{\rm Tr}}
\newcommand{\idol}{\ensuremath{\mathbbm 1}}
\def\Re{\mathop{\rm Re}}
\def\Im{\mathop{\rm Im}}

\usepackage{iopams}
\usepackage{graphicx}
\usepackage{bm}
\usepackage{bbm}
\usepackage{amsfonts}

\begin{document}

\title{Complete condition for nonzero quantum correlation in continuous variable systems}
\author{Chengjie Zhang$^{1,2}$, Sixia Yu$^{2,3}$, Qing Chen$^{2,4}$, Haidong Yuan$^{5}$, C.H. Lai$^{2,6}$,
C.H. Oh$^{2,6}$}
\address{$^{1}$College of Physics, Optoelectronics and Energy, Soochow University, Suzhou, 215006, China}
\address{$^{2}$Centre for quantum technologies, National University of Singapore, 3 Science Drive 2, Singapore 117543, Singapore}
\address{$^{3}$Hefei National Laboratory for Physical Sciences at Microscale and Department of Modern Physics,  University of Science and Technology of China, Hefei, Anhui 230026, China}
\address{$^{4}$Department of Physics, Yunnan University, Kunming, 650091, China}
\address{$^{5}$Department of Mechanical and Automation Engineering, The Chinese University of Hong Kong, Hong Kong}
\address{$^{6}$Physics department, National University of Singapore, 3 Science Drive 2, Singapore 117543, Singapore}
\date{\today}

\begin{abstract}
Quantum correlation provides a promising measure beyond entanglement. Here, we propose a necessary and sufficient condition for nonzero quantum correlation in continuous variable systems, which is simple and easy to perform in terms of a marker $Q_r$. In order to get this condition, we introduce continuous-variable local orthogonal bases of the operator space, which are generalized from the orthogonal basis sets in local operator space for discrete variables. Based on this, we obtain the marker $Q_r$ for all bipartite continuous variable states, and provide several examples including two-mode Gaussian and non-Gaussian states. Our result may provide a candidate for quantum correlation measures, and can be measured by designed quantum circuits.
\end{abstract}

Keywords: quantum correlation, continuous variables, quantum circuits
\section{Introduction}
In order to quantify the quantumness of correlations, quantum correlation has been proposed by Henderson and Vedral \cite{discord1} and independently by Ollivier and Zurek (they use the term ``quantum discord" instead of ``quantum correlation") \cite{discord2}. Interestingly, quantum correlation has found numerous applications \cite{RMP} in addition to its initial motivation in pointer states \cite{discord2}. For examples, relating to the difference between quantum and classical Maxwell's demons \cite{demon}, its role in open dynamics \cite{CP}, the local broadcasting of the quantum correlations \cite{broadcast,broadcast2}, the phase transitions \cite{phase}, its operational meaning in state merging protocols \cite{merging,merging2}, activating multipartite entanglement \cite{activate}, creating entanglement in the measurement process \cite{measure}, to be a resource in quantum state discrimination \cite{discrimination}, entanglement distribution \cite{distribution,distribution2}, remote state preparation \cite{remote}, observing the operational significance of quantum correlation consumption \cite{Gu}, guaranteeing a minimum precision in the optimal phase estimation protocol \cite{estimation}, interpretation as the difference in superdense coding capacities \cite{superdense}, and demonstrating that quantum discord cannot be shared \cite{share}.

It is widely accepted that quantum mutual information measures the total correlations in a system \cite{discord1} defined as
\begin{equation}\label{}
    I(\varrho_{AB})=S(\varrho_A)+S(\varrho_B)-S(\varrho_{AB}),
\end{equation}
where $S(\varrho)=-\mathrm{Tr}(\varrho\log_2\varrho)$ is the von Neumann entropy, and $\varrho_A$ ($\varrho_B$) is the reduced density matrix of subsystem $A$ ($B$). Classical correlation, which measures the maximal information gained from $\varrho_{AB}$
with measurement on one of the subsystems, is defined as
\begin{equation}\label{}
    C_A(\varrho_{AB})=\max_{\{E_k\}}[S(\varrho_B)-\sum_k p_k S(\varrho_{B|k})],
\end{equation}
where the maximum is taken over all possible positive operator-valued measures (POVM) \cite{discord1} $\{E_k\}$ on subsystem $A$  with $p_k=\mathrm{Tr}(E_k\otimes\idol\varrho_{AB})$ and $\varrho_{B|k}=\Trr_A(E_k\otimes\idol\varrho_{AB})/p_k$. Quantum correlation is therefore given by
\begin{equation}\label{}
    D_A(\varrho_{AB})=I(\varrho_{AB})-C_A(\varrho_{AB}).
\end{equation}
Alternatively, one can define $C_B(\varrho_{AB})$ and $D_B(\varrho_{AB})$ if the measurement is performed on subsystem $B$ instead of on $A$. In the following, we only consider the quantum correlation with measurement performing on subsystem $A$, since the quantum correlation on $B$ can be reduced to quantum correlation on $A$ by exchanging these two subsystems. A state $\varrho_{AB}$ is of zero quantum correlation with measurement performing on subsystem $A$ if and only if it can be written as
\begin{equation}\label{zerodiscord}
    \varrho_{AB}=\sum_i p_i |i\rangle\langle i|_A\otimes\varrho_B^i,
\end{equation}
where $\{|i\rangle_A\}$ is a set of orthonormal basis for subsystem $A$. Furthermore, the quantum correlation has been generalized to continuous variable systems using Gaussian measurements for two-mode Gaussian states \cite{gaussian1,gaussian2}, and several experiments have been done \cite{experiment1,experiment2,experiment3,experiment4,experiment5,experiment6,experiment7,experiment8}. It is worth noticing that non-discord like definitions of quantum correlations have been considered for continuous variable systems as well, related to their phase space representation, see Refs. \cite{Paris,Vogel} and references therein.

Because of the importance of quantum correlations, it becomes a fundamental issue to decide whether a given state contains nonzero quantum correlation or not for both discrete variables and continuous variables. For a given bipartite state in a finite-dimensional system, several conditions for nonzero quantum correlation have been proposed \cite{2N,Ferraro,condition1,condition2,condition3,condition4} in the form of local commutativity, strong positive partial transpose, etc. In particular, Daki\'{c}, Vedral and Brukner  \cite{condition1} have proposed a  necessary and sufficient condition for nonzero quantum correlation in bipartite finite-dimensional system. Subsequently, Refs. \cite{condition2,condition3} generalized the condition to continuous variable system. However, this condition in general requires calculating an infinite number of commutators for continuous variable states (since for an arbitrary $d\times d$ state there are $d^2(d^2-1)/2$ commutation relations to be checked), which cannot be implemented efficiently. Therefore, the necessary and sufficient condition for nonzero quantum correlation of bipartite continuous variable states still remains challenging.

In this paper, we derive a necessary and sufficient criterion for nonvanishing quantum correlation in continuous variable systems in terms of an expression $Q_r$ designated  as a marker. Unlike checking an infinite number of commutators as in Refs. \cite{condition1,condition2,condition3}, our criterion is possibly easier to test since $Q_r$ can be calculated directly from the Wigner function. In order to get the condition, we introduce the complete orthogonal basis sets in local Hilbert-Schmidt spaces of operators for continuous variables. Based on this condition, we obtain the marker $Q_r$ for all bipartite states, and present examples including all two-mode Gaussian states and several non-Gaussian states. Moreover, the criterion may provide a candidate for quantum correlation measures, and it can be efficiently measured by designed quantum circuits.

\section{Continuous-variable local operator basis}
In discrete variable systems, take a $d\times d$ system as an example, for each subsystem  a complete set of orthogonal bases of the operator space  $\{G_k\}$ consists of $d^2$ operators (which are not necessarily Hermitian) satisfying $\Trr(G_k^\dag G_l)=\delta_{kl}$ and $\varrho=\sum_{k=1}^{d^2}\Trr(\varrho G_k)G_k^\dag$. The local orthogonal operator basis has been widely used in many problems of discrete variables, such as entanglement detection \cite{YU}, necessary and sufficient condition for nonzero quantum correlation \cite{condition1}, and quantifying quantum uncertainty based on skew information \cite{luo3}.

Consider a two-mode continuous-variable state, for each subsystem a complete set of orthogonal bases of the operator space contains an infinite number of operators $G(\lambda)$ of this subsystem satisfying orthogonal relations
\begin{eqnarray}
\Trr[G^\dag(\lambda)G(\lambda')]=\delta^{(2)}(\lambda-\lambda'),
\end{eqnarray}
and complete-set condition $\varrho=\int\langle G(\lambda)\rangle_{\varrho}G^\dag(\lambda)\mathrm{d}^2\lambda$, where the index $\lambda$ is an arbitrary complex number, $\mathrm{d}^2\lambda:=\mathrm{d}\Re(\lambda)\mathrm{d}\Im(\lambda)$, and $\delta^{(2)}(\lambda-\lambda'):=\delta(\Re(\lambda-\lambda'))\delta(\Im(\lambda-\lambda'))$ with $\Re(z)$ and $\Im(z)$ being the real and imaginary part of the complex number $z$, respectively. All the integrals throughout the paper are from $-\infty$ to $+\infty$.
There are infinite complete sets of continuous-variable operator bases. For later use, we introduce a typical complete set of bases,
\begin{equation}\label{CVLOOs}
    \mathcal{G}(\lambda)=\frac{D(\lambda)}{\sqrt{\pi}},
\end{equation}
where $D(\lambda)$ is the Weyl displacement operator defined by $D(\lambda)=\exp(\lambda \hat{a}^\dag-\lambda^* \hat{a})$.

\section{Necessary and sufficient condition}
Consider a two-mode continuous-variable state $\varrho_{AB}$, let us choose the complete set of operator basis $\{\mathcal{G}(\lambda)\}$ in subsystem $B$. The density matrix $\varrho_{AB}$ of the two-mode system has a partial expression
\begin{eqnarray}
 \varrho_{AB}=\int \varrho^A(\lambda)\otimes\mathcal{G}^\dag(\lambda)\mathrm{d}^2\lambda,
\end{eqnarray}
with
\begin{equation}\label{}
 \varrho^A(\lambda)=\Trr_B[\varrho_{AB}\idol_A\otimes\mathcal{G}(\lambda)].
\end{equation}
From Eq. (\ref{zerodiscord}), one can show that the state $\varrho_{AB}$ is of zero quantum correlation if and only if there exists a von Neumann measurement $\{\Pi_i=|i\rangle\langle i|\}$ such that
\begin{equation}\label{}
    \sum_i(\Pi_i\otimes \idol_B)\varrho_{AB}(\Pi_i\otimes \idol_B)=\varrho_{AB}.
\end{equation}
Therefore, the necessary and sufficient condition becomes
\begin{equation}\label{}
    \int(\sum_i\Pi_i\varrho^A(\lambda)\Pi_i)\otimes\mathcal{G}^\dag(\lambda)\mathrm{d}^2\lambda=\int \varrho^A(\lambda)\otimes\mathcal{G}^\dag(\lambda)\mathrm{d}^2\lambda,
\end{equation}
which is equivalent to the set of infinite conditions:
\begin{equation}\label{}
    \sum_i\Pi_i\varrho^A(\lambda)\Pi_i=\varrho^A(\lambda)
\end{equation}
holds for arbitrary $\lambda$.
It means that the set of infinite number of operators $\{\varrho^A(\lambda)\}$ has a common eigenbasis defined by the set of projectors $\{\Pi_i\}$. Thus, the projector set  $\{\Pi_i\}$ exists if and only if
\begin{equation}\label{conditions}
    [\varrho^A(\lambda), \varrho^A(\lambda')]=0,  \   \  \  \  \  \  \forall \lambda,\lambda'.
\end{equation}
These infinite conditions can merge into a single condition: Eq. (\ref{conditions}) is satisfied if and only if the expression $Q=0$ where
\begin{eqnarray}
Q:=\frac{1}{2}\int\Trr\Big([\varrho^A(\lambda), \varrho^A(\lambda')][\varrho^A(\lambda), \varrho^A(\lambda')]^\dag\Big)\mathrm{d}^2\lambda\mathrm{d}^2\lambda',\label{8}
\end{eqnarray}
since on the one hand $[\varrho^A(\lambda), \varrho^A(\lambda')]=0$ for all possible $\lambda$ and $\lambda'$ leads directly to $Q=0$, and on the other hand if $Q=0$ then we have $[\varrho^A(\lambda), \varrho^A(\lambda')]=0$ because the integrand in the integral over $\lambda$ and $\lambda'$ in Eq. (\ref{8}) is nonnegative. \textit{Therefore, a state $\varrho_{AB}$ is of zero quantum correlation i.e., $D_A(\varrho_{AB})=0$ if and only if $Q=0$}.

For an arbitrary two-mode state $\varrho_{AB}$, its characteristic function is defined as the expectation value of the two-mode Weyl displacement operator
\begin{equation}\label{}
    \chi(\lambda_1,\lambda_2)=\Trr[\varrho_{AB} D_1(\lambda_1)D_2(\lambda_2)],
\end{equation}
and its Wigner function is defined as the Fourier transform of the characteristic
function
\begin{equation}\label{}
    W(\alpha_1,\alpha_2)=\frac{1}{\pi^{4}}\int \exp[\sum_{i=1}^{2}(\lambda_i^*\alpha_i-\lambda_i\alpha_i^*)]\chi(\lambda_1,\lambda_2)\mathrm{d}^2\lambda_1\mathrm{d}^2\lambda_2.
\end{equation}
Conversely, one can express $\varrho_{AB}$ using its Wigner function:
\begin{eqnarray}
    \varrho_{AB}=\frac{1}{\pi^{2}}\int W(\alpha_1,\alpha_2)\exp[-\sum_{i=1}^{2}\beta_i(\alpha_i^*-\hat{a}_i^{\dag})+\sum_{i=1}^{2}\beta_i^*(\alpha_i-\hat{a}_i)]\prod_{i=1}^{2}\mathrm{d}^2\alpha_i\mathrm{d}^2\beta_i.\nonumber
\end{eqnarray}
Based on this, we arrive at the expression of $Q$ using the Wigner function (see the Appendix):
\begin{equation}
  Q=Q_1-Q_2,
\end{equation}
with
\begin{eqnarray}
 Q_j&:=&4\pi^3\int q_j \exp[i4\Im(\alpha_3^*\alpha_1+\alpha_5^*\alpha_3-\alpha_5^*\alpha_1)]\prod_{i=1}^{5}\mathrm{d}^2\alpha_i,\\
 q_1&:=&W(\alpha_1,\alpha_2)W(\alpha_3,\alpha_2)W(\alpha_5,\alpha_4)W(\alpha_5-\alpha_3+\alpha_1,\alpha_4),\\
 q_2&:=&W(\alpha_1,\alpha_2)W(\alpha_3,\alpha_4)W(\alpha_5,\alpha_2)W(\alpha_5-\alpha_3+\alpha_1,\alpha_4),
\end{eqnarray}
and $j=1,2$. For later use, we introduce a ratio version of $Q$:
\begin{eqnarray}
Q_r=1-\frac{Q_2}{Q_1},\label{Qr2}
\end{eqnarray}
where $Q_r=0$ if and only if $Q=0$ since $Q_1>0$ holds for arbitrary states.
In order to decide whether a theoretically given two-mode state $\varrho_{AB}$ contains nonzero quantum correlation or not, one first gets its Wigner function based on the definition, and then calculates the integral in Eq. (\ref{Qr2}). After that one can easily check whether the integral result $Q_r$ is equal to zero or not. For an experimentally unknown state, one can measure $Q_r$ by using quantum circuits without any information of Wigner function, see the Appendix for details.

\section{Examples}
Let us define the position and momentum operators as $\hat{x}=(\hat{a}+\hat{a}^\dag)/2$ and $\hat{p}=-i(\hat{a}-\hat{a}^\dag)/2$, respectively.  The Wigner function of a zero-mean two-mode Gaussian state is \cite{cvrmp}:
\begin{equation}\label{}
    W(\xi)=\frac{1}{4\pi^{2}\sqrt{\mathrm{Det}\mathcal{V}}}\exp(-\xi \mathcal{V}^{-1}\xi^{\mathrm{T}}/2),
\end{equation}
where the four-dimensional vector $\xi$ has the quadrature pairs of all two-modes as its components $\xi=(x_1,p_1,x_2,p_2)$ with $\hat{\xi}=(\hat{x}_1,\hat{p}_1,\hat{x}_2,\hat{p}_2)$, and $\mathcal{V}$ is the covariance matrix defined by $\mathcal{V}_{ij}=\Trr[\varrho(\Delta\hat{\xi}_i\Delta\hat{\xi}_j+\Delta\hat{\xi}_j\Delta\hat{\xi}_i)/2]$ with $\Delta\hat{\xi}_i=\hat{\xi}_i-\langle\hat{\xi}_i\rangle$. There is a standard form for the covariance matrix $\mathcal{V}$ of two-mode Gaussian state,
\begin{eqnarray}
\mathcal{V}=\left(\begin{array}{cc}
\mathcal{A}& \mathcal{C}\\
\mathcal{C}^{T}& \mathcal{B}
  \end{array}\right)\;,
\end{eqnarray}
where $\mathcal{A}=\mathrm{diag}\{a,a\}$, $\mathcal{B}=\mathrm{diag}\{b,b\}$ and $\mathcal{C}=\mathrm{diag}\{c_1,c_2\}$ with $a,b\geq1/4$ and $ab\geq c_1^2,c_2^2$. Under this standard form, it can be figured out that (see the Appendix)
\begin{eqnarray}
Q_r=1-\frac{\sqrt{\prod_{i=1}^{2}[(ab-c_i^2)(b+16a^2b-16ac_i^2)]}}{a[b^2+16\prod_{i=1}^{2}(ab-c_i^2)]}.\label{Q3}
\end{eqnarray}
Therefore, for an arbitrary two-mode Gaussian state, it is easy to check whether or not it contains nonzero quantum correlation using Eq. (\ref{Q3}).

As an example of two-mode Gaussian states, let us consider the covariance matrix of two-mode symmetrical squeezed thermal state given by
\begin{eqnarray}
a=b=[(1+2n)\cosh2r]/4,\\
c_1=-c_2=[(1+2n)\sinh2r]/4,
\end{eqnarray}
with  $r$ and $n$ being the squeezing parameter and average photon number, respectively. For $n=0$, it is reduced to a pure two-mode squeezed vacuum state. Using Eq. (\ref{Q3}), one can derive
\begin{equation}\label{}
   Q_r=\frac{\sinh^22r}{\sinh^22r+1+(1+2n)^2}
\end{equation}
for the two-mode squeezed thermal states.
The covariance matrix of two-mode squeezed thermal state has the property $c_1=-c_2$. Actually, for all the two-mode Gaussian states with covariance matrix satisfying
\begin{equation}\label{}
    c_1^2=c_2^2\equiv c^2,
\end{equation}
the expression of $Q_r$  in Eq. (\ref{Q3}) can be simplified as
\begin{equation}\label{}
    Q_r=\frac{bc^2}{A+bc^2}
\end{equation}
with $A=(ab-c^2)[b+16a(ab-c^2)]$. Therefore, in this case $Q_r=0$ if and only if $c=0$ since $b\geq1/4$, or in the limit $A/(bc^2)\rightarrow\infty$ we have $Q_r\rightarrow0$. For instance, for the two-mode symmetrical squeezed thermal state, it has zero quantum correlation if and only if (i) $r=0$ or (ii) $Q_r\rightarrow0$ when $n\rightarrow\infty$ or $r\rightarrow0$.

When will Eq. (\ref{Q3}) be equal to zero for a general two-mode Gaussian state? As discussed above, there are two situations: (i) $Q_r$ is exactly equal to zero; (ii) $Q_r\rightarrow0$ in the limit. For the first situation (i), in order to get $Q_r=0$ one needs to check whether
\begin{eqnarray}
 f&:=&\prod_{i=1}^{2}[(ab-\chi_i)(b+16a^2b-16a\chi_i)]-a^2[b^2+16\prod_{i=1}^{2}(ab-\chi_i)]^2\nonumber\\
 &=&16ab(g-ab^3/16)
\end{eqnarray}
is equal to zero, where we have denoted
\begin{eqnarray}
 \chi_i:=c_i^2, \\
 g:=(2ab-\chi_1-\chi_2-\Delta)\prod_{i=1}^{2}(ab-\chi_i),\\
 \Delta:=b(32a^2-1)/(16a).
\end{eqnarray}
There are two cases:
\begin{itemize}
\item[(a)] If $g\leq0$ then $f\leq-a^2b^4<0$.
\item[(b)] If $g>0$, then $\partial g/\partial \chi_1=-(ab-\chi_2)(3ab-2\chi_1-\chi_2-\Delta)\leq0$ and $\partial g/\partial \chi_2=-(ab-\chi_1)(3ab-\chi_1-2\chi_2-\Delta)\leq0$. It means $g$ is a monotonic decreasing function of $\chi_1$ and $\chi_2$. The maximal value of  $g$ is $ab^3/16$ when $\chi_1=\chi_2=0$. Thus, $f\leq0$ holds.
\end{itemize}
Therefore, it can be concluded that $f\leq0$ holds for both cases and $f=0$ if and only if $\chi_1=\chi_2=0$, i.e., $c_1=c_2=0$. For the second situation (ii), it can be seen that if $\prod_{i=1}^{2}[(ab-c_i^2)(b+16a^2b-16ac_i^2)]^{1/2}/[ab^2+16a\prod_{i=1}^{2}(ab-c_i^2)]\rightarrow1$ then $Q_r\rightarrow0$. It is worth noticing that Ref. \cite{gaussian1} has obtained the result of the first situation. The difference is that the optimization of Ref. \cite{gaussian1} is just over all Gaussian measurements, but the optimization of our method is over all possible POVM measurements which are much more general than Gaussian measurements. Our result of the first situation coincides with the results given by Refs. \cite{condition2,Optimality} which considered generic measurements for Gaussian states.

In general, one can also use Eq. (\ref{Qr2}) to detect the quantum correlation of an arbitrary two-mode non-Gaussian state. Since non-Gaussian states have no standard form as Gaussian states, we show several examples to demonstrate how to detect quantum correlation of non-Gaussian states based on Eq. (\ref{Qr2}) (see the Appendix).

As the first example of non-Gaussian states, consider the photon-number mixed state
\begin{eqnarray}
\varrho=k|00\rangle\langle00|+\bar{k}|+1\rangle\langle+1|
\end{eqnarray}
with $\bar{k}=1-k$, $|+\rangle=(|0\rangle+|1\rangle)/\sqrt{2}$ and $0\leq k\leq1$. This state is a simple non-Gaussian state in a finite dimension. Based on the definition of Wigner function, its Wigner function is
\begin{eqnarray}
W(\alpha_1,\alpha_2)=W_{0}(\alpha_1,\alpha_2)\big[k+2\bar{k}(x_1+|\alpha_1|^2)(4|\alpha_2|^2-1)\big],\nonumber
\end{eqnarray}
with $W_{0}(\alpha_1,\alpha_2)=4\exp[-2(|\alpha_1|^2+|\alpha_2|^2)]/\pi^2$ being Wigner function of $|00\rangle$, $\alpha_i=x_i+i p_i$ and $x_i$, $p_i$ are real parameters. After some algebra, one can get
\begin{eqnarray}
 Q_r=\frac{k^2\bar{k}^2}{2(k^4+k^2\bar{k}^2+\bar{k}^4)}
\end{eqnarray}
using Eq. (\ref{Qr2}). Therefore, this photon-number mixed state is of zero-correlation if and only if $Q_r=0$, i.e., $k=0$ or $1$. On the other hand, the quantum correlation for subsystem $B$ of $\varrho$, $D_B(\varrho)$ is always zero when $0\leq k\leq1$. Actually, this finite-dimensional state can also be detected by the condition shown in Ref. \cite{condition1}. We present it here to show that our necessary and sufficient condition is not only suitable for continuous-variable system but also for finite-dimensional system.

The second example of two-mode non-Gaussian state is a convex combination of vacuum state and a zero-mean Gaussian state with covariance matrix in the standard form satisfying $c_1^2=c_2^2\equiv c^2$, i.e.,
\begin{eqnarray}
 \varrho=k\varrho_{G}+\bar{k}|00\rangle\langle00|
\end{eqnarray}
with $0\leq k\leq1$. $\varrho_{G}$ is the two-mode Gaussian state satisfying $c_1^2=c_2^2$ with its Wigner function having been previously introduced, and $|00\rangle$ is vacuum state with the Wigner function being $W_{0}(\alpha_1,\alpha_2)$. In general, the convex combination of two Gaussian states may not be a Gaussian state. Based on Eq. (\ref{Qr2}), one can figure out that
\begin{eqnarray}
Q_r=\frac{k^4bc^2}{32(A+bc^2)AQ_1}+\frac{8k^2\bar{k}^2 c^2}{(B+4c^2)BQ_1}+ \frac{128  \bar{k} k^3  (1 + 4 b) c^2}{(C+8 c^2 + 32 b c^2)CQ_1}
\end{eqnarray}
with
\begin{eqnarray}
A=(ab-c^2)[b+16a(ab-c^2)],\\
B=(1 + 4 a)^2 b - 8 (1 + 2 a) c^2,\\
C=[(1 + 4 a) (1 + 4 b) - 16 c^2]B.
\end{eqnarray}
Therefore, the  Gaussian state satisfying $c_1^2=c_2^2$ with vacuum-state noise contains zero quantum correlation if and only if (i) $k=0$ or (ii) $c=0$ or (iii) in the limit $A/(bc^2)\rightarrow\infty$ when $k=1$ or (iv) in the limit all the three terms $A/(k^4bc^2),B/(k^2\bar{k}^2c^2),C/[\bar{k} k^3(1+4b)c^2]\rightarrow\infty$. For example, if the Gaussian state $\varrho_{G}$ is the two-mode symmetrical squeezed thermal state, then $Q_r=0$ if and only if (i) $k=0$ or (ii) $r=0$ or (iii) $Q_r\rightarrow0$ when $n\rightarrow\infty$ or  $r\rightarrow0$ or $k\rightarrow0$.

\begin{figure}[tbp]\centering
\includegraphics[scale=0.4]{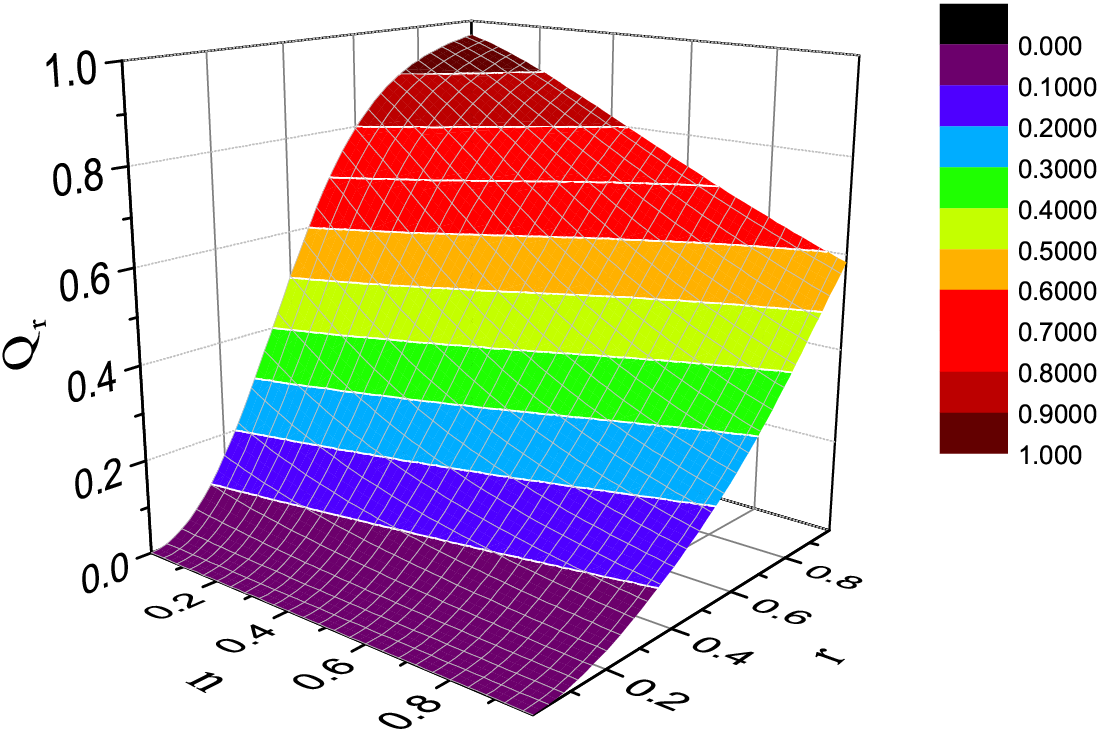}
\caption{$Q_r$ versus parameters $n$ and $r$ for the single-photon-added two-mode symmetrical squeezed thermal states. All the states in the range except the ones with $r=0$ can be detected as nonzero quantum correlation states by the necessary and sufficient condition.}\label{fig1}
\end{figure}

As the last example, let us consider the single-photon-added two-mode symmetrical squeezed thermal state with its Wigner function given by
\begin{eqnarray}
W(\alpha_1,\alpha_2)&=&\frac{W_{STS}(\alpha_1,\alpha_2)}{(1 + 2 n)^2 (\cosh^2r + n \cosh2r)}\nonumber\\
&&\times\Big[(x_2 + 2 n x_2 + x_2\cosh2r-x_1 \sinh2r)^2 \nonumber\\
&&+ (p_2 + 2 n p_2 + p_2 \cosh2r + p_1 \sinh2r)^2\nonumber\\
&&-(1 + 2 n) (n + \cosh^2r)\Big],\nonumber
\end{eqnarray}
where $\alpha_i=x_i+i p_i$ and $x_i$, $p_i$ are real parameters, and
$W_{STS}(\alpha_1,\alpha_2)=4[(1 + 2 n) \pi]^{-2}\exp[-2(|\alpha_1|^2+|\alpha_2|^2) \cosh2r/(1 + 2 n) + 4(x_1 x_2 - p_1 p_2) \sinh2r/(1 + 2 n)]$
is the Wigner function of two-mode symmetrical squeezed thermal state. When $n=0$, this state is reduced to the single-photon-added two-mode squeezed vacuum state shown in Ref. \cite{singleadd}. Using Eq. (\ref{Qr2}), we have checked all the single-photon-added two-mode symmetrical squeezed thermal states in the range of $0\leq n\leq1$ and $0\leq r\leq1$ where the results have been shown in Fig. \ref{fig1}. All the states except the ones with $r=0$ in Fig. \ref{fig1} can be detected as nonzero quantum correlation states by our necessary and sufficient condition: $D_A(\varrho_{AB})=0$ if and only if $Q_r=0$, i.e., $r=0$ in Fig. \ref{fig1}. Furthermore, we also find that $Q_r\rightarrow0$ when $n\rightarrow\infty$ or  $r\rightarrow0$.

\section{Measuring $Q_r$ for an unknown state by quantum circuits}
  Here we shall show that the marker $Q_r$ in Eq. (\ref{Qr2}) can be equivalently written as Eq. (\ref{Qr3}).  We start from Eq. (\ref{8}):
\begin{eqnarray}
Q&=&\frac{1}{2}\int\Trr\Big([\varrho^A(\lambda), \varrho^A(\lambda')][\varrho^A(\lambda), \varrho^A(\lambda')]^\dag\Big)\mathrm{d}^2\lambda\mathrm{d}^2\lambda'\nonumber\\
&=&\int\Trr\Big(\varrho^A(\lambda)\varrho^A(\lambda')\varrho^A(\lambda')^\dag\varrho^A(\lambda)^\dag\Big)\mathrm{d}^2\lambda\mathrm{d}^2\lambda'\nonumber\\
&&-\int\Trr\Big(\varrho^A(\lambda)\varrho^A(\lambda')\varrho^A(\lambda)^\dag\varrho^A(\lambda')^\dag\Big)\mathrm{d}^2\lambda\mathrm{d}^2\lambda'.
\end{eqnarray}
It is worth noticing that $\varrho^A(\lambda)=\Trr_B[\varrho_{AB}\idol_A\otimes\mathcal{G}(\lambda)]$. We introduce continuous variable version of swap operator $V=\sum_{i,j=0}^{\infty}|ij\rangle\langle ji|=\int \mathcal{G}(\lambda)\otimes \mathcal{G}^\dag(\lambda)\mathrm{d}^2\lambda$, and 4-cyclic permutation operator $X=\sum_{i,j,k,l=0}^{\infty}|ijkl\rangle\langle jkli|=V_{12}V_{23}V_{34}$. The swap operator has the properties $V|\phi_1\rangle|\phi_2\rangle=|\phi_2\rangle|\phi_1\rangle$ and $\Trr (V A\otimes B)=\Trr(AB)$. Similarly, the  4-cyclic permutation operator $X$ has properties $X|\phi_1\rangle|\phi_2\rangle|\phi_3\rangle|\phi_4\rangle=|\phi_4\rangle|\phi_1\rangle|\phi_2\rangle|\phi_3\rangle$ and $\Trr(X A\otimes B\otimes C\otimes D)=\Trr(DCBA)$. Therefore,
\begin{eqnarray}
Q_1&=&\int\Trr\Big(\varrho^A(\lambda)\varrho^A(\lambda')\varrho^A(\lambda')^\dag\varrho^A(\lambda)^\dag\Big)\mathrm{d}^2\lambda\mathrm{d}^2\lambda'\nonumber\\
&=&\int\Trr\Big(X^A\varrho^A(\lambda)^\dag\otimes\varrho^A(\lambda')^\dag\otimes\varrho^A(\lambda')\otimes\varrho^A(\lambda)\Big)\mathrm{d}^2\lambda\mathrm{d}^2\lambda'\nonumber\\
&=&\int\Trr\Big(X^A\mathcal{G}^\dag(\lambda)\otimes\mathcal{G}^\dag(\lambda')\otimes\mathcal{G}(\lambda')\otimes\mathcal{G}(\lambda)\varrho_{AB}^{\otimes4}\Big)\mathrm{d}^2\lambda\mathrm{d}^2\lambda'\nonumber\\
&=&\Trr\Big(X^AV_{14}^B V_{23}^B\varrho_{AB}^{\otimes4}\Big),
\end{eqnarray}
where we have used property of swap operator $V=V^\dag=\int \mathcal{G}^\dag(\lambda)\otimes \mathcal{G}(\lambda)\mathrm{d}^2\lambda$, and property of 4-cyclic permutation operator $\Trr(X A\otimes B\otimes C\otimes D)=\Trr(DCBA)$. Similarly, we have
\begin{eqnarray}
Q_2&=&\int\Trr\Big(\varrho^A(\lambda)\varrho^A(\lambda')\varrho^A(\lambda)^\dag\varrho^A(\lambda')^\dag\Big)\mathrm{d}^2\lambda\mathrm{d}^2\lambda'\nonumber\\
&=&\int\Trr\Big(X^A\varrho^A(\lambda')^\dag\otimes\varrho^A(\lambda)^\dag\otimes\varrho^A(\lambda')\otimes\varrho^A(\lambda)\Big)\mathrm{d}^2\lambda\mathrm{d}^2\lambda'\nonumber\\
&=&\int\Trr\Big(X^A\mathcal{G}^\dag(\lambda')\otimes\mathcal{G}^\dag(\lambda)\otimes\mathcal{G}(\lambda')\otimes\mathcal{G}(\lambda)\varrho_{AB}^{\otimes4}\Big)\mathrm{d}^2\lambda\mathrm{d}^2\lambda'\nonumber\\
&=&\Trr\Big(X^AV_{13}^B V_{24}^B\varrho_{AB}^{\otimes4}\Big).
\end{eqnarray}
Therefore,
\begin{eqnarray}
&&Q=Q_1-Q_2=\Trr\Big(X^AV_{14}^B V_{23}^B\varrho_{AB}^{\otimes4}\Big)-\Trr\Big(X^AV_{13}^B V_{24}^B\varrho_{AB}^{\otimes4}\Big),\\
&&Q_r=1-\frac{Q_2}{Q_1}=1-\frac{\Trr\Big(X^AV_{13}^B V_{24}^B\varrho_{AB}^{\otimes4}\Big)}{\Trr\Big(X^AV_{14}^B V_{23}^B\varrho_{AB}^{\otimes4}\Big)}=1-\frac{\mathrm{Tr}(U_2\varrho_{AB}^{\otimes4})}{\mathrm{Tr}(U_1\varrho_{AB}^{\otimes4})},\label{Qr3}
\end{eqnarray}
where $U_1=X^A V_{14}^B V_{23}^B$ and $U_2=X^AV_{13}^B V_{24}^B$ with $Q_i=\mathrm{Tr}(U_i\varrho_{AB}^{\otimes4})$ ($i=1,2$).

\begin{figure}[tbp]\centering
\includegraphics[scale=0.58]{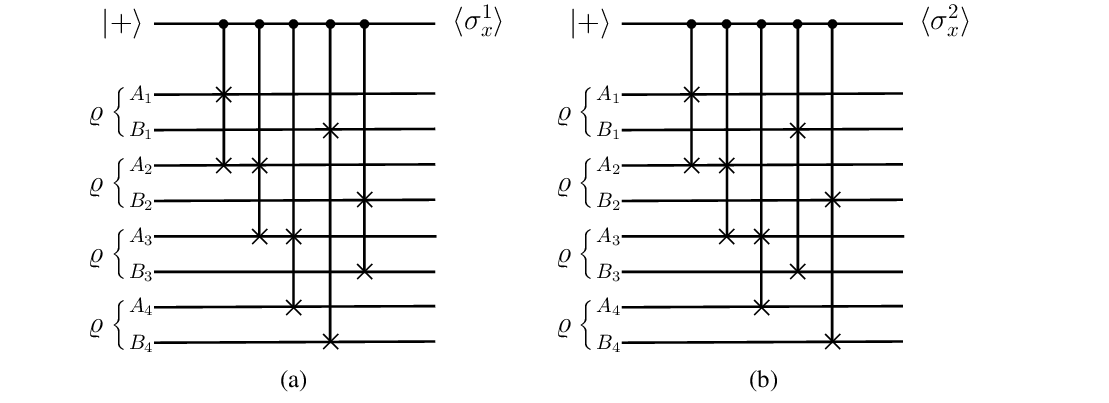}
\caption{Quantum circuits to measure the marker $Q_r$. Each of the two quantum circuits has one auxiliary qubit initially prepared in the state $|+\rangle$, and four copies of initial state $\varrho_{AB}$ are needed. Each auxiliary qubit is the source of five controlled-swap gates represented by connected crosses. After the controlled-swap operations, we perform $\sigma_x$ measurements on the two auxiliary qubits: (a) shows the measurement for $U_1$ and (b) corresponds to the measurement for $U_2$, with results being $\langle\sigma_x^i\rangle=\mathrm{Tr}(U_i\varrho_{AB}^{\otimes4})$ ($i=1,2$).}\label{fig2}
\end{figure}

In order to measure $Q_r$ for an experimentally unknown state, we have designed  quantum circuits as shown in Fig. \ref{fig2} which are generalized from Ref. \cite{yu2} to continuous variable systems. Each of the two quantum circuits has one auxiliary qubit initially prepared in the state $|+\rangle=(|0\rangle+|1\rangle)/\sqrt{2}$, and four copies of initial state $\varrho_{AB}$ are needed. After several controlled-swap operations with each auxiliary qubit as a source, we perform $\sigma_x$ measurements on the two auxiliary qubits. Take $U_1$ as an example, the total state including the auxiliary qubit is
\begin{equation}\label{}
    \varrho_{total}=\frac{1}{2}\sum_{i,i'=0}^{1}\Big(|i\rangle\langle i'|\otimes U_1^i\varrho_{AB}^{\otimes4}{U_1^{i'}}^\dag\Big).
\end{equation}
Thus, we have
\begin{equation}\label{}
  \langle\sigma_x^1\rangle=\frac{1}{2}[\Trr(\varrho_{AB}^{\otimes4}U_1^\dag)+\Trr(U_1\varrho_{AB}^{\otimes4})]=\Trr(U_1\varrho_{AB}^{\otimes4}).
\end{equation}
Similarly, $\langle\sigma_x^2\rangle=\mathrm{Tr}(U_2\varrho_{AB}^{\otimes4})$ holds for $U_2$. Therefore, according to Eq. (9), one has
\begin{equation}\label{}
  Q_r=1-\frac{\langle \sigma_x^2\rangle}{\langle \sigma_x^1\rangle}.
\end{equation}
It is worth noticing that the measurements do not need any information about the Wigner function of the experimentally unknown state $\varrho_{AB}$.

One possible physical setup in experiments is the trapped-ion system. Wang proposed a physical scheme to realize the controlled-swap gate in a trapped-ion system \cite{wang}, which is the only kind of quantum gate employed in our quantum circuits. However, it may be not easy to realize the whole quantum circuits under current experimental technique, since we simultaneously need four copies of target state.

\section{Discussion and conclusion}
From all the examples of Gaussian states and non-Gaussian states above, one can see that in principle the two terms in the integrand of Eq. (\ref{Qr2}) can always be analytically calculated, since for Gaussian states they are Gaussian integrals like $\int\exp(-ax^2+bx+c)\mathrm{d}x$ with $a>0$, and for non-Gaussian states they can be expressed as integrals of Gaussian function with a polynomial factor like $\int\exp(-ax^2+bx+c)f(x)\mathrm{d}x$ where $f(x)=\sum_{k=0}^{n}a_kx^k$ and $a>0$. Based on the identities $\int_{-\infty}^{+\infty}\exp(-ax^2+bx+c)\mathrm{d}x=\sqrt{\pi/a}\exp[b^2/(4a) + c] $ and $\int_{-\infty}^{+\infty} x^n\exp(-ax^2+bx+c)\mathrm{d}x= \sqrt{\pi/a}\exp[b^2/(4a) + c]\sum_{k=0}^{\lfloor n/2\rfloor}n!b^{n-2k}(2a)^{k-n}/[2^k k!(n-2k)!]$ where $a>0$, $n$ is a positive integer and the floor function $\lfloor x\rfloor$ is the largest integer not greater than $x$, the analytical results of all the integrals can be obtained. On the other hand, the integral Eq. (\ref{Qr2}) can always be calculated since it relates to the expectation value of the quantum correlation witness shown in Ref. \cite{yu2}, which is always a real number.

In fact, $Q_r$ is non-negative for all the two-mode states, since the integrand in the integral over $\lambda$ and $\lambda'$ in Eq. (\ref{8}) is nonnegative and $Q_1>0$, and for two-mode squeezed vacuum states when the squeezing parameter $r\rightarrow\infty$ (i.e., the original, normalized Einstein-Podolsky-Rosen state) the marker $Q_r$ reaches its limit $Q_r\rightarrow1$. From this point of view, $Q_r$ may be regarded as a candidate for quantum correlation measures, because it is always nonnegative, and $Q_r=0$ if and only if the state is of zero quantum correlation. If a zero quantum correlation state $\sigma_0$ has experimental imperfections, for example, it mixes with an arbitrarily small amount of noise, i.e., $\sigma_0'=(\sigma_0+\epsilon \varrho)/(1+\epsilon)$ where $\epsilon$ is an arbitrarily small quantity and $\varrho$ is a nonzero-correlation state. One can see that $Q_r(\sigma_0')\sim O(\epsilon)$, since from Eq. (\ref{Qr3}) we have
\begin{eqnarray}
Q_r(\sigma_0')=1-\frac{\mathrm{Tr}(U_2\sigma_0'^{\otimes4})}{\mathrm{Tr}(U_1\sigma_0'^{\otimes4})}&\sim& \frac{[\mathrm{Tr}(\Delta U\sigma_0^{\otimes4})+\epsilon\mathrm{Tr}(\Delta U\varpi)+O(\epsilon^2)]/(1+\epsilon)^4}{\mathrm{Tr}(U_1\sigma_0^{\otimes4})}\nonumber\\
&\sim&\epsilon\frac{\mathrm{Tr}(\Delta U\varpi)}{\mathrm{Tr}(U_1\sigma_0^{\otimes4})}+O(\epsilon^2)\nonumber\\
&\sim&O(\epsilon),
\end{eqnarray}
where $\epsilon\rightarrow0^+$, $\Delta U=U_1-U_2$, $\varpi=\varrho\otimes\sigma_0^{\otimes3}+\sigma_0\otimes\varrho\otimes\sigma_0^{\otimes2}+\sigma_0^{\otimes2}\otimes\varrho\otimes\sigma_0+\sigma_0^{\otimes3}\otimes\varrho$, and $\mathrm{Tr}(\Delta U\sigma_0^{\otimes4})=0$ because $\sigma_0$ is a zero quantum correlation state. That is to say $Q_r(\sigma_0')$ is also arbitrarily small, which means $\sigma_0'$ has arbitrarily small amount of quantum correlation. For general cases, both quantum correlation and $Q_r$ are continuous functions of $\epsilon$ which determines the experimental imperfect, and when $\epsilon=0$ the quantum correlation and $Q_r$ of the experimental imperfect state should be zero. Thus, when $\epsilon$ is arbitrarily small, the experimental imperfect state has arbitrarily small amount of quantum correlation and $Q_r$, and small $Q_r$ does imply that the experimental imperfect state has very small quantum correlation. In that sense, the marker $Q_r$ is robust with respect to experimental imperfections.

Last but not least, the set of zero correlation states is of measure zero and a randomly chosen state will never have zero correlation \cite{Ferraro}, but this does not mean that the condition (\ref{Qr2}) is useless. In fact, there are infinitely many zero correlation states, and it is a fundamental but not an easy problem to decide whether a theoretically given state contains zero quantum correlation or not. A necessary and sufficient condition for zero correlation in discrete variable system has been found \cite{condition1}, but for continuous variables the necessary and sufficient condition was missing until condition (\ref{Qr2}) is proposed.

To summarize, we have derived a necessary and sufficient criterion of nonvanishing quantum correlation for arbitrary two-mode states in continuous variable systems in terms of a marker $Q_r$. This criterion is simple and easy to perform without checking an infinite number of commutators. We have introduced the complete orthogonal basis sets in local Hilbert-Schmidt spaces of operators for continuous variables. The criterion may provide a candidate for quantum correlation measures, and can be efficiently measured by designed quantum circuits.

\section*{Acknowledgements}
This work is funded by the Singapore Ministry of Education (partly through the Academic Research Fund Tier 3 MOE2012-T3-1-009), the National Research Foundation, Singapore (Grant No. WBS: R-710-000-008-271), the financial support from RGC of Hong Kong(Grant No. 538213), and the National Natural Science Foundation of China (Grant No. 11075227).

\section*{Appendix}
Recall that we choose the complete set of orthogonal basis for operator space as $\mathcal{G}(\lambda)=D(\lambda)/\sqrt{\pi}$. Using the identities
\begin{eqnarray}
\Trr [D(\lambda)]=\pi\delta^{(2)}(\lambda),\\
D^\dag(\lambda)=D(-\lambda),\\
D(\lambda_1)D(\lambda_2)=D(\lambda_1+\lambda_2)\exp[(\lambda_1\lambda_2^*-\lambda_1^*\lambda_2)/2],
\end{eqnarray}
one can easily check the orthogonal relation $\Trr[\mathcal{G}^\dag(\lambda)\mathcal{G}(\lambda')]=\delta^{(2)}(\lambda-\lambda')$ for Eq. (\ref{CVLOOs}).

In order to obtain Eq. (\ref{Qr2}), let us notice that the characteristic function is defined as the expectation value of the two-mode Weyl displacement operator
\begin{eqnarray}
\chi(\lambda_1,\lambda_2)=\Trr[\varrho_{AB} D_1(\lambda_1)D_2(\lambda_2)],
\end{eqnarray}
and its Wigner function is defined as the Fourier transform of the characteristic function
\begin{eqnarray}
W(\alpha_1,\alpha_2)=\frac{1}{\pi^{4}}\int \exp[\sum_{i=1}^{2}(\lambda_i^*\alpha_i-\lambda_i\alpha_i^*)]\chi(\lambda_1,\lambda_2)\mathrm{d}^2\lambda_1\mathrm{d}^2\lambda_2.
\end{eqnarray}
Therefore,
\begin{eqnarray}
\varrho_{AB}&=&\frac{1}{\pi^2}\int\Trr[\varrho_{AB}D_1^\dag(\lambda_1)\otimes D_2^\dag(\lambda_2)]D_1(\lambda_1)\otimes D_2(\lambda_2)\mathrm{d}^2\lambda_1\mathrm{d}^2\lambda_2,\nonumber\\
&=&\frac{1}{\pi^2}\int\chi(-\lambda_1,-\lambda_2)D_1(\lambda_1)\otimes D_2(\lambda_2)\mathrm{d}^2\lambda_1\mathrm{d}^2\lambda_2,
\end{eqnarray}
\begin{eqnarray}
\chi(-\lambda_1,-\lambda_2)=\int\exp[\sum_{i=1}^{2}(\lambda_i^*\alpha_i-\lambda_i\alpha_i^*)]W(\alpha_1,\alpha_2)\mathrm{d}^2\alpha_1\mathrm{d}^2\alpha_2.
\end{eqnarray}
Combining this two equations, we have
\begin{eqnarray}
\varrho_{AB}&=&\frac{1}{\pi^2}\int W(\alpha_1,\alpha_2)\exp[\sum_{i=1}^{2}(\lambda_i^*\alpha_i-\lambda_i\alpha_i^*)]D_1(\lambda_1)\otimes D_2(\lambda_2)\prod_{i=1}^{2}\mathrm{d}^2\alpha_i\mathrm{d}^2\lambda_i,\nonumber\\
&=&\frac{1}{\pi^{2}}\int W(\alpha_1,\alpha_2)\exp[-\sum_{i=1}^{2}\lambda_i(\alpha_i^*-\hat{a}_i^{\dag})+\sum_{i=1}^{2}\lambda_i^*(\alpha_i-\hat{a}_i)]\nonumber\\
&&\times\prod_{i=1}^{2}\mathrm{d}^2\alpha_i\mathrm{d}^2\lambda_i.
\end{eqnarray}
Using the definition $\varrho^A(\lambda):=\Trr_B[\varrho_{AB}\mathcal{G}(\lambda)]$, it can be concluded that
\begin{eqnarray}
\varrho^A(\lambda)&=&\frac{1}{\pi^{\frac{5}{2}}}\int W(\alpha_1,\alpha_2)\exp[-\beta_1(\alpha_1^*-\hat{a}_1^{\dag})+\beta_1^*(\alpha_1-\hat{a}_1)]\nonumber\\
&&\times\Trr\{\exp[-\beta_2(\alpha_2^*-\hat{a}_2^{\dag})+\beta_2^*(\alpha_2-\hat{a}_2)]D(\lambda)\}\prod_{i=1}^{2}\mathrm{d}^2\alpha_i\mathrm{d}^2\beta_i\nonumber\\
&=&\frac{1}{\pi^{\frac{5}{2}}}\int W(\alpha_1,\alpha_2)\exp[-\beta_1(\alpha_1^*-\hat{a}_1^{\dag})+\beta_1^*(\alpha_1-\hat{a}_1)]\exp(-\beta_2\alpha_2^*+\beta_2^*\alpha_2)\nonumber\\
&&\times\Trr[D(\beta_2)D(\lambda)]\prod_{i=1}^{2}\mathrm{d}^2\alpha_i\mathrm{d}^2\beta_i\nonumber\\
&=&\frac{1}{\pi^{\frac{5}{2}}}\int W(\alpha_1,\alpha_2)\exp[-\beta_1(\alpha_1^*-\hat{a}_1^{\dag})+\beta_1^*(\alpha_1-\hat{a}_1)]\exp(-\beta_2\alpha_2^*+\beta_2^*\alpha_2)\nonumber\\
&&\times\pi\delta^{(2)}(\beta_2+\lambda)\prod_{i=1}^{2}\mathrm{d}^2\alpha_i\mathrm{d}^2\beta_i\nonumber\\
&=&\frac{1}{\pi^{\frac{3}{2}}}\int W(\alpha_1,\alpha_2)\exp(-\beta_1\alpha_1^*+\beta_1^*\alpha_1)\exp(\lambda\alpha_2^*-\lambda^*\alpha_2)\nonumber\\
&&\times D(\beta_1)\mathrm{d}^2\alpha_1\mathrm{d}^2\alpha_2\mathrm{d}^2\beta_1,
\end{eqnarray}
where we have used the identities $\Trr[D(\beta_2)D(\lambda)]=\pi\delta^{(2)}(\beta_2+\lambda)$ and $\int f(\beta_2)\delta^{(2)}(\beta_2+\lambda)\mathrm{d}^2\beta_2=f(-\lambda)$.
Based on the expression of $\varrho^A(\lambda)$, $Q_1$ can be written as
\begin{eqnarray}
Q_1&=&\int\Trr\Big(\varrho^A(\lambda)\varrho^A(\lambda')\varrho^A(\lambda')^\dag\varrho^A(\lambda)^\dag\Big)\mathrm{d}^2\lambda\mathrm{d}^2\lambda'\nonumber\\
&=&\frac{1}{\pi^6}\int W(\alpha_1,\alpha_2)\exp(-\beta_1\alpha_1^*+\beta_1^*\alpha_1)\exp(\lambda\alpha_2^*-\lambda^*\alpha_2)
W(\alpha_3,\alpha_4)\nonumber\\
&&\times W(\alpha_5,\alpha_6)\exp(\beta_5\alpha_5^*-\beta_5^*\alpha_5)\exp(-\lambda'\alpha_6^*+\lambda'^*\alpha_6)
W(\alpha_7,\alpha_8)\nonumber\\
&&\times\exp(-\beta_3\alpha_3^*+\beta_3^*\alpha_3)\exp(\lambda'\alpha_4^*-\lambda'^*\alpha_4)\exp(\beta_7\alpha_7^*-\beta_7^*\alpha_7)\exp(-\lambda\alpha_8^*+\lambda^*\alpha_8)\nonumber\\
&&\times \Trr[D(\beta_1)D(\beta_3)D^\dag(\beta_5)D^\dag(\beta_7)]\mathrm{d}^2\lambda\mathrm{d}^2\lambda'\mathrm{d}^2\beta_1\mathrm{d}^2\beta_3\mathrm{d}^2\beta_5\mathrm{d}^2\beta_7\prod_{i=1}^{8}\mathrm{d}^2\alpha_i.
\end{eqnarray}
It is worth noticing that $\Trr[D(\beta_1)D(\beta_3)D^\dag(\beta_5)D^\dag(\beta_7)]=\pi\delta^{(2)}(\beta_1+\beta_3-\beta_5-\beta_7)\exp[(\beta_1\beta_3^*-\beta_1^*\beta_3)/2]\exp\{[-(\beta_1+\beta_3)\beta_5^*+(\beta_1^*+\beta_3^*)\beta_5]/2\}$, $\int \exp[\lambda(\alpha_2^*-\alpha_8^*)-\lambda^*(\alpha_2-\alpha_8)]\mathrm{d}^2\lambda=\pi^2\delta^{(2)}(\alpha_2-\alpha_8)$ and $\int \exp[\lambda'(\alpha_4^*-\alpha_6^*)-\lambda'^*(\alpha_4-\alpha_6)]\mathrm{d}^2\lambda'=\pi^2\delta^{(2)}(\alpha_4-\alpha_6)$. Thus, integrating with respect to $\lambda$, $\lambda'$, $\beta_7$, $\alpha_6$ and $\alpha_8$, we have
\begin{eqnarray}
Q_1&=&\int\Trr\Big(\varrho^A(\lambda)\varrho^A(\lambda')\varrho^A(\lambda')^\dag\varrho^A(\lambda)^\dag\Big)\mathrm{d}^2\lambda\mathrm{d}^2\lambda'\nonumber\\
&=&\frac{1}{\pi}\int W(\alpha_1,\alpha_2)\exp(-\beta_1\alpha_1^*+\beta_1^*\alpha_1)W(\alpha_3,\alpha_4)\exp(-\beta_3\alpha_3^*+\beta_3^*\alpha_3)\nonumber\\
&&\times W(\alpha_5,\alpha_6)\exp(\beta_5\alpha_5^*-\beta_5^*\alpha_5) W(\alpha_7,\alpha_8)\exp(\beta_7\alpha_7^*-\beta_7^*\alpha_7)\nonumber\\
&&\times\exp[(\beta_1\beta_3^*-\beta_1^*\beta_3)/2]\exp\{[-(\beta_1+\beta_3)\beta_5^*+(\beta_1^*+\beta_3^*)\beta_5]/2\}\nonumber\\
&&\times \delta^{(2)}(\beta_1+\beta_3-\beta_5-\beta_7)\delta^{(2)}(\alpha_2-\alpha_8)\delta^{(2)}(\alpha_4-\alpha_6)\mathrm{d}^2\beta_1\mathrm{d}^2\beta_3\mathrm{d}^2\beta_5\mathrm{d}^2\beta_7\prod_{i=1}^{8}\mathrm{d}^2\alpha_i\nonumber\\
&=&\frac{1}{\pi}\int W(\alpha_1,\alpha_2)\exp(-\beta_1\alpha_1^*+\beta_1^*\alpha_1)W(\alpha_3,\alpha_4)\exp(-\beta_3\alpha_3^*+\beta_3^*\alpha_3)\nonumber\\
&&\times W(\alpha_5,\alpha_4)\exp(\beta_5\alpha_5^*-\beta_5^*\alpha_5) W(\alpha_7,\alpha_2)\nonumber\\
&&\times\exp[(\beta_1+\beta_3-\beta_5)\alpha_7^*-(\beta_1^*+\beta_3^*-\beta_5^*)\alpha_7]\exp[(\beta_1\beta_3^*-\beta_1^*\beta_3)/2]\nonumber\\
&&\times\exp\{[-(\beta_1+\beta_3)\beta_5^*+(\beta_1^*+\beta_3^*)\beta_5]/2\}\mathrm{d}^2\beta_1\mathrm{d}^2\beta_3\mathrm{d}^2\beta_5\mathrm{d}^2\alpha_7\prod_{i=1}^{5}\mathrm{d}^2\alpha_i.
\end{eqnarray}
Using the identity $\int\exp[\beta_5(\alpha_5^*-\alpha_7^*+\frac{\beta_1^*+\beta_3^*}{2})-\beta_5^*(\alpha_5-\alpha_7+\frac{\beta_1+\beta_3}{2})]\mathrm{d}^2\beta_5=\pi^2\delta^{(2)}(\alpha_5-\alpha_7+\frac{\beta_1+\beta_3}{2})$, we have,
\begin{eqnarray}
Q_1&=&\int\Trr\Big(\varrho^A(\lambda)\varrho^A(\lambda')\varrho^A(\lambda')^\dag\varrho^A(\lambda)^\dag\Big)\mathrm{d}^2\lambda\mathrm{d}^2\lambda'\nonumber\\
&=&\pi\int W(\alpha_1,\alpha_2)\exp(-\beta_1\alpha_1^*+\beta_1^*\alpha_1)W(\alpha_3,\alpha_4)\exp(-\beta_3\alpha_3^*+\beta_3^*\alpha_3)\nonumber\\
&&\times W(\alpha_5,\alpha_4)W(\alpha_7,\alpha_2)
\exp[(\beta_1\beta_3^*-\beta_1^*\beta_3)/2]\exp[(\beta_1+\beta_3)\alpha_7^*-(\beta_1^*+\beta_3^*)\alpha_7]\nonumber\\
&&\times\delta^{(2)}(\alpha_5-\alpha_7+\frac{\beta_1+\beta_3}{2})\mathrm{d}^2\beta_1\mathrm{d}^2\beta_3\mathrm{d}^2\alpha_7\prod_{i=1}^{5}\mathrm{d}^2\alpha_i.
\end{eqnarray}
Notice that $\delta^{(2)}(\alpha_5-\alpha_7+\frac{\beta_1+\beta_3}{2})=4\delta^{(2)}(2\alpha_5-2\alpha_7+\beta_1+\beta_3)$ and $\int\exp[\beta_1(-\alpha_1^*+\alpha_3^*+\alpha_7^*-\alpha_5^*)-\beta_1^*(-\alpha_1+\alpha_3+\alpha_7-\alpha_5)]\mathrm{d}^2\beta_1=\pi^2\delta^{(2)}(-\alpha_1+\alpha_3+\alpha_7-\alpha_5)$, one can integrate over $\beta_3$, $\beta_1$ and $\alpha_3$,
\begin{eqnarray}
Q_1&=&\int\Trr\Big(\varrho^A(\lambda)\varrho^A(\lambda')\varrho^A(\lambda')^\dag\varrho^A(\lambda)^\dag\Big)\mathrm{d}^2\lambda\mathrm{d}^2\lambda'\nonumber\\
&=&4\pi\int W(\alpha_1,\alpha_2)W(\alpha_3,\alpha_4)W(\alpha_5,\alpha_4)W(\alpha_7,\alpha_2)\nonumber\\
&&\times\exp[\beta_1(-\alpha_1^*+\alpha_3^*+\alpha_7^*-\alpha_5^*)-\beta_1^*(-\alpha_1+\alpha_3+\alpha_7-\alpha_5)]\nonumber\\
&&\times\exp[i4\Im(\alpha_3\alpha_7^*-\alpha_3\alpha_5^*-\alpha_5\alpha_7^*)]\mathrm{d}^2\beta_1\mathrm{d}^2\alpha_7\prod_{i=1}^{5}\mathrm{d}^2\alpha_i\nonumber\\
&=&4\pi^3\int W(\alpha_1,\alpha_2)W(\alpha_3,\alpha_4)W(\alpha_5,\alpha_4)W(\alpha_7,\alpha_2)\delta^{(2)}(-\alpha_1+\alpha_3+\alpha_7-\alpha_5)\nonumber\\
&&\times\exp[i4\Im(\alpha_3\alpha_7^*-\alpha_3\alpha_5^*-\alpha_5\alpha_7^*)]\mathrm{d}^2\alpha_7\prod_{i=1}^{5}\mathrm{d}^2\alpha_i\nonumber\\
&=&4\pi^3\int W(\alpha_1,\alpha_2)W(\alpha_7,\alpha_2)W(\alpha_5,\alpha_4)W(\alpha_1+\alpha_5-\alpha_7,\alpha_4)\nonumber\\
&&\times\exp[i4\Im(\alpha_1\alpha_7^*-\alpha_1\alpha_5^*+\alpha_5^*\alpha_7)]\mathrm{d}^2\alpha_1\mathrm{d}^2\alpha_2\mathrm{d}^2\alpha_4\mathrm{d}^2\alpha_5\mathrm{d}^2\alpha_7.\nonumber
\end{eqnarray}
Exchange the indices $\alpha_3$ and $\alpha_7$, it can be obtained that
\begin{eqnarray}
Q_1&=&\int\Trr\Big(\varrho^A(\lambda)\varrho^A(\lambda')\varrho^A(\lambda')^\dag\varrho^A(\lambda)^\dag\Big)\mathrm{d}^2\lambda\mathrm{d}^2\lambda'\nonumber\\
&=&4\pi^3\int W(\alpha_1,\alpha_2)W(\alpha_3,\alpha_2)W(\alpha_5,\alpha_4)W(\alpha_1+\alpha_5-\alpha_3,\alpha_4)\nonumber\\
&&\times\exp[i4\Im(\alpha_1\alpha_3^*-\alpha_1\alpha_5^*+\alpha_5^*\alpha_3)]\prod_{i=1}^{5}\mathrm{d}^2\alpha_i.
\end{eqnarray}
Similarly, we can also obtain that
\begin{eqnarray}
Q_2&=&\int\Trr\Big(\varrho^A(\lambda)\varrho^A(\lambda')\varrho^A(\lambda)^\dag\varrho^A(\lambda')^\dag\Big)\mathrm{d}^2\lambda\mathrm{d}^2\lambda'\nonumber\\
&=&4\pi^3\int W(\alpha_1,\alpha_2)W(\alpha_3,\alpha_4)W(\alpha_5,\alpha_2)W(\alpha_5-\alpha_3+\alpha_1,\alpha_4)\nonumber\\
&&\times\exp[i4\Im(\alpha_3^*\alpha_1+\alpha_5^*\alpha_3-\alpha_5^*\alpha_1)]\prod_{i=1}^{5}\mathrm{d}^2\alpha_i.
\end{eqnarray}
Therefore, Eq. (\ref{Qr2}) holds for arbitrary two-mode state. In order to decide whether a theoretically given two-mode state $\varrho_{AB}$ contains nonzero quantum discord or not, one first obtains its Wigner function based on the definition, and then calculates the integral in Eq. (\ref{Qr2}) using the obtained Wigner function. After that one can easily check whether the integral result $Q_r$ is equal to zero or not. For an experimentally unknown state, one can measure $Q_r$ by using quantum circuits without any information of Wigner function.

For two-mode Gaussian states, we define the position and momentum operators as $\hat{x}=(\hat{a}+\hat{a}^\dag)/2$ and $\hat{p}=-i(\hat{a}-\hat{a}^\dag)/2$, respectively.  The Wigner function of the two-mode Gaussian state is \cite{cvrmp}:
\begin{equation}\label{}
    W(\alpha_1,\alpha_2)=\frac{1}{4\pi^2\sqrt{\mathrm{Det}\mathcal{V}}}\exp\Big(-\frac{1}{2}\xi \mathcal{V}^{-1}\xi^{\mathrm{T}}\Big),
\end{equation}
where the four-dimensional vector $\xi=(x_1,p_1,x_2,p_2)$ and $\alpha_i=x_i+ip_i$. Substituting the Gaussian state Wigner function into Eq. (\ref{Qr2}), and using the Gaussian function integral formula ten times:
\begin{eqnarray}
\int_{-\infty}^{+\infty}\exp(-ax^2+bx+c)\mathrm{d}x=\sqrt{\frac{\pi}{a}}\exp\Bigg[\frac{b^2}{4a} + c\Bigg],  \ \ \ \ \ \mathrm{with}\ \  a>0,
\end{eqnarray}
one can finally obtain Eq. (\ref{Q3}).

For two-mode non-Gaussian states, in general, we first obtain the Wigner function of the non-Gaussian state based on the definition, and then calculate the integral in Eq. (\ref{Qr2}) using its Wigner function. One can see that in principle the two integral terms in Eq. (\ref{Qr2}) can always be analytically calculated since they can be expressed as integrals of Gaussian function with a polynomial factor like $\int\exp(-ax^2+bx+c)f(x)\mathrm{d}x$ where $f(x)=\sum_{k=0}^{n}a_kx^k$ and $a>0$, which we only need to use the formula
\begin{eqnarray}
\int_{-\infty}^{+\infty} x^n\exp(-ax^2+bx+c)\mathrm{d}x=\sqrt{\frac{\pi}{a}}\exp\Bigg[\frac{b^2}{4a} + c\Bigg]\sum_{k=0}^{\lfloor n/2\rfloor}\frac{n!b^{n-2k}(2a)^{k-n}}{2^k k!(n-2k)!}, \nonumber
\end{eqnarray}
where $n$ is a positive integer, $a>0$, and the floor function $\lfloor x\rfloor$ is the largest integer not greater than $x$.


\section*{References}
\begin{thebibliography}{99}
\bibitem{discord1} Henderson L and Vedral V 2001 \textit{J. Phys. A} \textbf{34}, 6899-6905

\bibitem{discord2} Ollivier H and Zurek W H 2001 \textit{Phys. Rev. Lett.} \textbf{88}, 017901

\bibitem{RMP} Modi K, Brodutch A, Cable H, Paterek T and Vedral V 2012 \textit{Rev. Mod. Phys.} \textbf{84}, 1655

\bibitem{demon} Zurek W H 2003 \textit{Phys. Rev. A} \textbf{67}, 012320

\bibitem{CP} Rodr\'{\i}guez-Rosario C A \textit{et al.} 2008 \textit{J. Phys. A} \textbf{41}, 205301

\bibitem{broadcast} Piani M, Horodecki P and Horodecki R 2008 \textit{Phys. Rev. Lett.} \textbf{100}, 090502

\bibitem{broadcast2} Luo S and Sun W 2010 \textit{Phys. Rev. A} \textbf{82}, 012338

\bibitem{phase} Werlang T, Trippe C, Ribeiro G A P and Rigolin G 2010 \textit{Phys. Rev. Lett.} \textbf{105}, 095702

\bibitem{merging} Madhok V and Datta A 2011 \textit{Phys. Rev. A} \textbf{83}, 032323

\bibitem{merging2} Cavalcanti D \textit{et al.} 2011 \textit{Phys. Rev. A} \textbf{83}, 032324

\bibitem{activate} Piani M \textit{et al.} 2011 \textit{Phys. Rev. Lett.} \textbf{106}, 220403

\bibitem{measure} Streltsov A, Kampermann H and Bru\ss D 2011 \textit{Phys. Rev. Lett.} \textbf{106}, 160401

\bibitem{discrimination} Roa L, Retamal J C and Alid-Vaccarezza M 2011 \textit{Phys. Rev. Lett. } \textbf{107}, 080401

\bibitem{distribution} Streltsov A, Kampermann H and Bru\ss D 2012 \textit{Phys. Rev. Lett.} \textbf{108}, 250501

\bibitem{distribution2} Chuan T K \textit{et al.} 2012 \textit{Phys. Rev. Lett.} \textbf{109}, 070501

\bibitem{remote} Daki\'{c} B \textit{et al.} 2012 \textit{Nat. Phys.} \textbf{8}, 666

\bibitem{Gu} Gu M \textit{et al.} 2012 \textit{Nat. Phys.} \textbf{8}, 671

\bibitem{estimation} Girolami D, Tufarelli T and Adesso G 2013 \textit{Phys. Rev. Lett.} \textbf{110}, 240402

\bibitem{superdense} Meznaric S, Clark S R and Datta A 2013 \textit{Phys. Rev. Lett.} \textbf{110}, 070502

\bibitem{share} Streltsov A and Zurek W H 2013 \textit{Phys. Rev. Lett.} \textbf{111}, 040401

\bibitem{gaussian1} Adesso G and Datta A 2010 \textit{Phys. Rev. Lett.} \textbf{105}, 030501

\bibitem{gaussian2} Giorda P and Paris M G A 2010 \textit{Phys. Rev. Lett.} \textbf{105}, 020503

\bibitem{experiment1} Madsen L S, Berni A, Lassen M and Andersen U L 2012 \textit{Phys. Rev. Lett.} \textbf{109}, 030402

\bibitem{experiment2} Blandino R \textit{et al.} 2012 \textit{Phys. Rev. Lett.} \textbf{109}, 180402

\bibitem{experiment3} Auccaise R \textit{et al.} 2011 \textit{Phys. Rev. Lett.} \textbf{107}, 070501

\bibitem{experiment4} Aguilar G H \textit{et al.} 2011 \textit{Phys. Rev. Lett.} \textbf{108}, 063601

\bibitem{experiment5} Silva I A \textit{et al.} 2013 \textit{Phys. Rev. Lett.} \textbf{110}, 140501

\bibitem{experiment6} Lanyon B P \textit{et al.} 2013 \textit{Phys. Rev. Lett.} \textbf{111}, 100504

\bibitem{experiment7} Xu J S \textit{et al.} 2010 \textit{Nat. Commun.} \textbf{1}, 7

\bibitem{experiment8} Xu J S \textit{et al.} 2013 \textit{Nat. Commun.} \textbf{4}, 2851

\bibitem{Paris} Ferraro A and Paris M G A 2012 \textit{Phys. Rev. Lett.} \textbf{108}, 260403

\bibitem{Vogel} Agudelo E, Sperling J, Vogel W 2013 \textit{Phys. Rev. A} \textbf{87}, 033811

\bibitem{2N} Bylicka B and Chru\'{s}ci\'{n}ski D 2010 \textit{Phys. Rev. A} \textbf{81}, 062102

\bibitem{Ferraro} Ferraro A, Aolita L, Cavalcanti D, Cucchietti F M and Ac\'{\i}n A 2010 \textit{Phys. Rev. A} \textbf{81}, 052318

\bibitem{condition1} Daki\'{c} B, Vedral V and Brukner \v{C} 2010 \textit{Phys. Rev. Lett.} \textbf{105}, 190502

\bibitem{condition2} Rahimi-Keshari S, Caves C M and Ralph T C 2013 \textit{Phys. Rev. A} \textbf{87}, 012119

\bibitem{condition3} Hosseini S \textit{et al.} 2014 \textit{J. Phys. B} \textbf{47}, 025503

\bibitem{condition4} Chen L, Chitambar E, Modi K and Vacanti G 2011 \textit{Phys. Rev. A} \textbf{83}, 020101(R)

\bibitem{YU} Yu S and Liu N L 2005 \textit{Phys. Rev. Lett.} \textbf{95}, 150504

\bibitem{luo3} Luo S L 2006 \textit{Phys. Rev. A} \textbf{73}, 022324

\bibitem{cvrmp} Braunstein S L and van Loock P 2005 \textit{Rev. Mod. Phys.} \textbf{77}, 513

\bibitem{Optimality} Pirandola S \textit{et al.} 2014 \textit{Phys. Rev. Lett.} \textbf{113}, 140405

\bibitem{singleadd} Agarwal G S  2011 \textit{New J. Phys.} \textbf{13}, 073008

\bibitem{yu2} Yu S, Zhang C, Chen Q and Oh C H 2011 arXiv:1102.4710

\bibitem{wang} Wang X 2001 \textit{J. Phys. A} \textbf{34}, 9577
\end{thebibliography}
\end{document}